\documentstyle[12pt,epsf]{article}
\title{
Thermodynamics of pseudospin-electron model
in mean field approximation}
\author{I.V.Stasyuk, O.D.Danyliv \\
\it   Institute for Condensed Matter Physics\\
\it   1~Svientsitskii str., Lviv 290011, Ukraine\\
\it tel./fax: + 380 (322) 761978
}
\date{6/10/1999}

\begin{document}
\maketitle


\begin{abstract}
The mean field type approach based on the self-consistent consideration of an
effective field created by electron transfer is developed for a
description of thermodynamics of the Hubbard type models with an infinitely
large
on-site repulsion. This procedure, formulated according to \cite{Izyumov94_J},
is an extension of the recently proposed generalized random phase
approximation(GRPA).

Within this approach, the thermodynamic properties of the two-sublattice
pseudo\-spin-electron model (the Hubbard model with local anharmonicity) are
studied.  Such a model can be used for a description of dielectric properties
of YBaCuO-type superconductors along $c$-axis; pseudospins represent
anharmonic motions of apical oxygens O4. It is shown that there are either
phase transitions in the model with jumps of the mean values of a pseudospin
and  of electron concentration (in the $\mu=const$ regime) or the  phase
separation (in the $n=const$ regime). The phase transitions or phase
separation are caused by pseudospin-pseudospin interaction as well as by
electron transfer (the latter results in appearing of effective interaction
between pseudospins).  The possibility of the ferroelectric type ordering of
pseudospins is investigated.

\end{abstract}

\section{Introduction}
The Hubbard model \cite{Hubbard63} is a fundamental one in the theory of
strongly correlated electron systems.  After the discovery of high-temperature superconductors
it has been intensively studied, because the HTSC-compounds necessarily
possess planes with strong correlation  between electrons described  by this
model. Amongst the most laborious but promising approaches to study the
Hubbard model there is a diagrammatic expansion with respect to atomic limit,
which essentially differs
from the diagrammatic technique for Fermi-systems \cite{Abrikosov75}.
One of recently worked out developments
%of the Hubbard-I approximation
is the GRPA proposed by Izyumov et.al. in order to study
the Hubbard model with Coulomb repulsion $U=\infty$ \cite{Izyumov92_I}-\cite{Izyumov92_J}. Application
of the GRPA approach to a pseudospin-electron system can be found in
\cite{SS93}-\cite{SSD95}. However, this approach, being developed for investigation of
correlation functions, is not a self-consistent theory. Therefore, lattice
instabilities  within it are treated as a result of singularities of
susceptibilities that cannot suffice when the model undergoes the
first order phase transition. The question how to calculate the
thermodynamic quantities within the GRPA remains open as well.

Returning to the possible approaches to a description of models with
strong electron correlation, one should note that due to a success in
obtaining exact results for simple models  (Falicov-Kimball model
\cite{Letfulov98_E1},\cite{Letfulov98_E2},
pseudospin-electron model with $U=0$ \cite{SS99}, etc), the approximation
of the infinite dimension of space
becomes very popular. However, there are some problems here, too. The
single-site problem for the Hubbard model remains analytically unsolved; there are some
difficulties with obtaining the higher order expansions in powers of
$1/d$.

In this paper we use the mean field approximation, proposed in \cite{Izyumov94_J} for
investigation of the $t-J$ model. We show using as an example the Hubbard model
that the diagrammatic series, summed over in this approximation expands
the diagrammatic series of the GRPA, and, hence, we can talk about the
self-consistent approach within the GRPA. The approximation is
applied to the two-sublattice pseudospin-electron model which, being derived
for description of ferroelectric type effects in HTSC of  YBaCuO type (see
\cite{Ruani94}-\cite{Conradson89}),
takes into account the two-sublattice structure of an apex oxygen subsystem in
such superconductors. Coupling between the electron and pseudospin
subsystems ( pseudospin variables are used to describe
the local strongly anharmonic vibrations
of apex oxygens) is performed via an interaction of
the $g n_i S_i^z$ type proposed  by M\"uller \cite{r4}.

The paper is composed as follows. In Section  2 the method of
construction of mean field approximation within the GRPA scheme
for the Hubbard model with
$U=\infty$ is presented. Application of the method to the two-sublattice
pseudospin-electron model is given in Section 3. Sections 4 and 5 contain
the results of numeric calculations and conclusions, respectively.

\section{MFA approximation applied for Hubbard model with infinite
Coulomb repulsion
}

{\bf In this section we show how to obtain the MFA equations in example case of
one-band Hubbard model and discuss the accuracy of that approach.}
The Hamiltonian of the Hubbard model without taking into account the
influence of magnetic field reads

\begin{equation}
H=-\mu\sum\limits_{i,s} n_{i}^{s}+U\sum\limits_{i} n_{i}^{\uparrow}
n_{i}^{\downarrow} +\sum\limits_{i,j,s} t_{ij} a_{is}^{+} a_{js},
\label{eq1}
\end{equation}
where $\mu$ is the chemical potential;
 $U$ is a Coulomb repulsion of electrons on the same orbital; $t_{ij}$
describes an electron transfer from the site $j$ to the site $i$
of a real lattice. $a_{is}$ is the annihilation operator of an
electron with spin $s$ ($s$ takes two values -- $\uparrow$ and
$\downarrow$), $n_{i}^{s}=a_{is}^{+}a_{is}$ is the electron number
operator.

Usually near the atomic limit the model
is considered using the basis of four states  $\mid
R_{i}\rangle\equiv\mid
n_{i}^{\uparrow},n_{i}^{\downarrow}\rangle$
\begin{equation}
\begin{array}{l}
\mid 1\rangle=\mid 0,0\rangle \\
\mid 2\rangle=\mid 1,1\rangle \\
\mid 3\rangle=\mid 0,1\rangle \\
\mid 4\rangle=\mid 1,0\rangle
\end{array}
\end{equation}
{\bf and Hubbard operators which are defined as follows}
$X_{i}^{RS}=\mid
R_{i}\rangle\langle
S_{i}\mid$.

Let us introduce the ``bare'' (zero-order) Matsubara Green's function
as \cite{Stas74}
\begin{eqnarray}
\begin{array}{l}
g_{ij}^{pq}(\tau-\tau')=\delta_{ij}\frac{\langle TX_{i}^{pq}(\tau)
X_{j}^{qp}(\tau')\rangle_{0}}{\langle\left[X_{i}^{pq},X_{j}^{qp}\right]_{\pm}
\rangle_{0}}= \delta_{ij}{\rm e}^{\varepsilon^{pq}(\tau-\tau')}\left\{
\begin{array}{l}
\pm n_{\pm}(\varepsilon^{pq}), \quad \tau>\tau'\\
\pm n_{\pm}(\varepsilon^{pq})-1, \quad \tau<\tau',
\end{array}
\right.\\
n_{\pm}(\varepsilon^{pq})=\frac{1}{{\rm e}^{\beta\varepsilon^{pq}}\pm 1},
\quad \varepsilon^{pq}=\varepsilon^{p}-\varepsilon^{q}
\end{array}
\end{eqnarray}
{\bf where $T$ means $T$-product,
$X_{i}^{pq}(\tau)={\rm e}^{\tau H_{0}}X_{i}^{pq}{\rm e}^{-\tau H_{0}}$
is an operator in
interaction representation and $\langle\ldots\rangle_{0}$ is the average
with the Gibbs distribution with the zero-order Hamiltonian.
$\varepsilon^{p}$ are the single-site energy levels of the zero-order
Hamiltonian
 (at  $t_{ij}=0$):}
\begin{equation}
\varepsilon^{1}=0, \quad \varepsilon^{2}=-2\mu+U,
\varepsilon^{3}=\varepsilon^{4}=-\mu.
\end{equation}
The upper sign is chosen when the operators $X_{i}^{pq}$ are of the
Fermi-type $(X^{41}$, $X^{23}$, $X^{31}$, $X^{24}$ and complex conjugate);
otherwise, the lower sign should be taken. Correspondingly, [~...~]$_{+}$
is an anticommutator, and [~...~]$_{-}$ is a commutator.

{\bf The results of perturbation theory and applying of Wick's thorem
\cite{Stas74,Izyumov89}
can be presented by elements
which are denoted in the following way}
$$
\begin{array}{l}
g_{ij}^{pq}(\tau-\tau')=\epsfxsize=3em \epsfbox{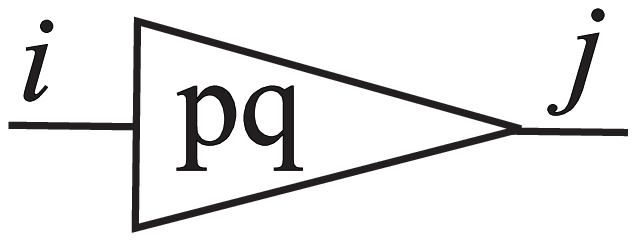}\\
\langle X_{i}^{pp}\rangle_{0}=\epsfxsize=1em \epsfbox{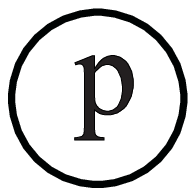},
\quad \langle X_{i}^{pp}X_{i}^{qq}\rangle_{0}^{c}\equiv
\frac{\partial}{\partial(-\beta\varepsilon^{p})} \langle
X_{i}^{qq}\rangle_{0}= \epsfxsize=3.5em \epsfbox{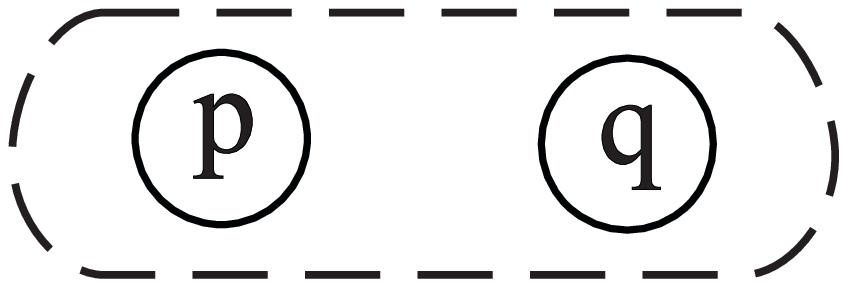}
\\
\langle X_{i}^{pp}X_{i}^{qq}X_{i}^{mm}\rangle_{0}^{c}\equiv \frac{\partial}
{\partial(-\beta\varepsilon^{p})}
\frac{\partial}{\partial(-\beta\varepsilon^{q})} \langle X_{i}^{mm}
\rangle_{0}= \epsfxsize=5em \epsfbox{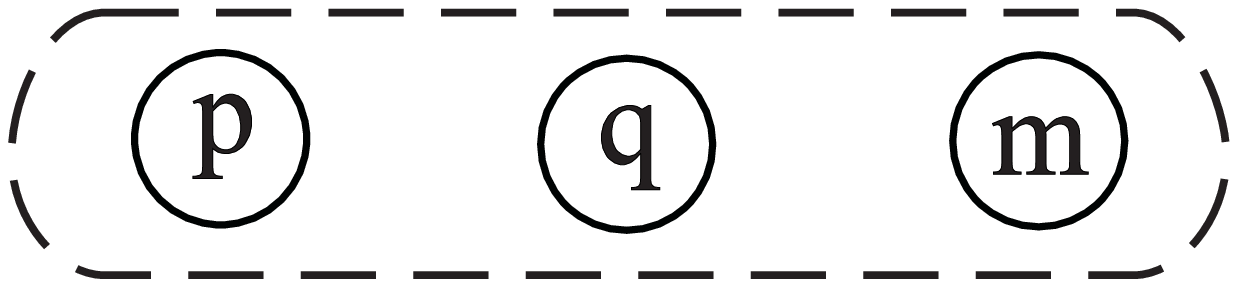} \\
g_{ij}^{pq}(\tau-\tau')\langle X_{i}^{pp}+X_{i}^{qq}\rangle
\equiv g_{ij}^{pq}(\tau-\tau')B_{i}^{pq}= \epsfxsize=3em \epsfbox{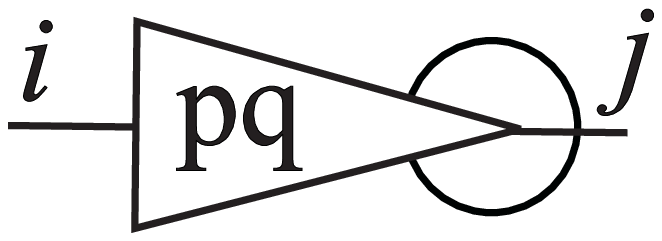} \\
t_{ij}= \epsfxsize=3em \epsfbox{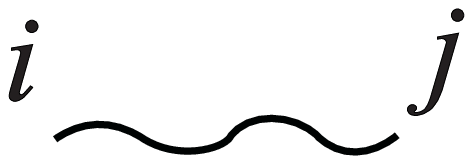} \\
\end{array}
$$

{\bf Let us write here the diagrammatic expansion for the average of
$X$-operator in Heisenberg representation
$\langle T\tilde{X}_i^{11} (\tau)\rangle=
\langle T{\rm e}^{\tau H} X_{i}^{11} {\rm e}^{-\tau H} \rangle$,
$i=1$: }

\begin{equation}
\label{eq10b}
\epsfxsize=14cm \epsfbox{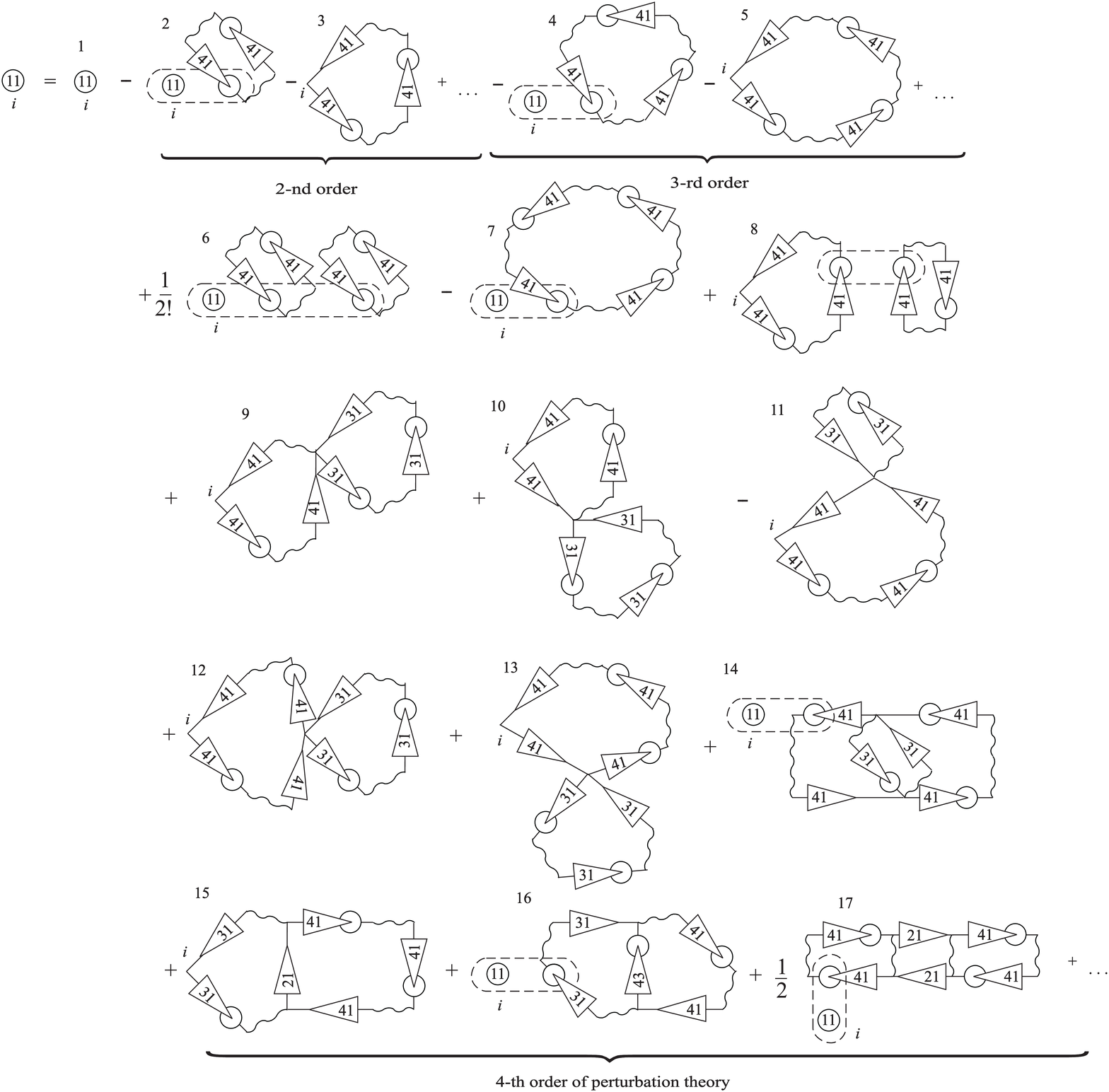}
\end{equation}

Retaining the diagrams with only simple loops, we can neglect
the diagrams which contain the Green's functions $g_{ij}^{43}(\tau-\tau')$. In
the limiting case $U=\infty$, which we shall consider hereafter, we can
neglect the Fermi-type Green's functions $g_{ij}^{24}$, $g_{ij}^{23}$ and the
Bose-type Green's function $g_{ij}^{21}$ as well. Discarding the
mentioned class of diagrams, one can show that the results of the application of Wick's
theorem do not depend on the so-called priority rules
\cite{Izyumov92_P}.

Classifying the obtained result, we treat the diagrams 2, 4, 6, 7 as a
``thickening'' of the zero-order average $\langle
X^{11}\rangle_{0}$ by the insets of the mean field type. The three
last diagrams in \ref{eq10b} should be discarded from the  reasons presented
above.There are also other two important type of diagrams: the diagrams
9, 10 renormalize the vertex parts of the Green's function of the initial
diagram 3, whereas the diagrams like 11, 12, 13
can be interpreted as an complication of self-energy parts in the more simple
diagram 3. After Fourier transformation all of the mentioned
diagrams (unlike the diagrams
14-17) correspond to the one sum (or to a certain power of the one sum)
over the wavevector {\bf k} and frequency $\omega$. As one can see from
the figure, the diagrams 3, 5, 9-13  are those which are summed
over within the GRPA when an equation of the Bete-Salpeter type is
solved \cite{Izyumov92_P},\cite{SSD95}.

Let us introduce a set of diagrams corresponding to the
generalized Hubbard-I approximation (in the simplified version that is
valid in the case of independent subbands \cite{SS94}), denoting it
by a ``thickened'' Green's function.
\begin{equation}
\epsfxsize=8cm \epsfbox{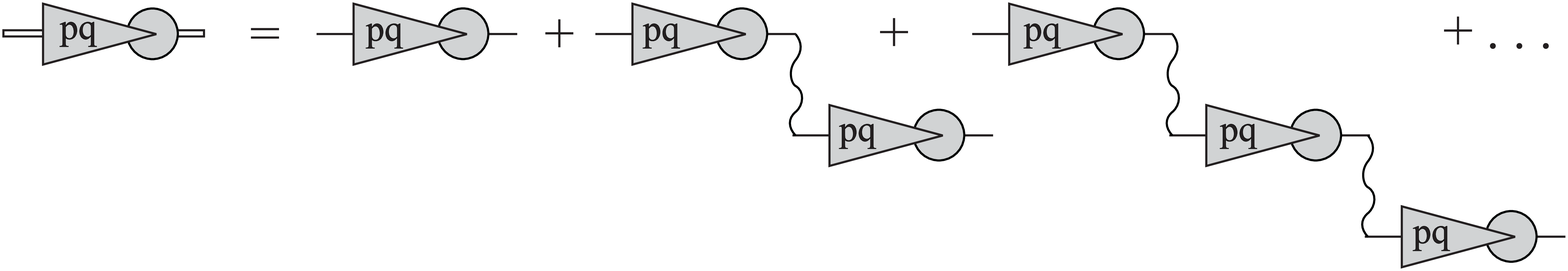} ,
\end{equation}
where $ \epsfxsize=1cm \epsfbox{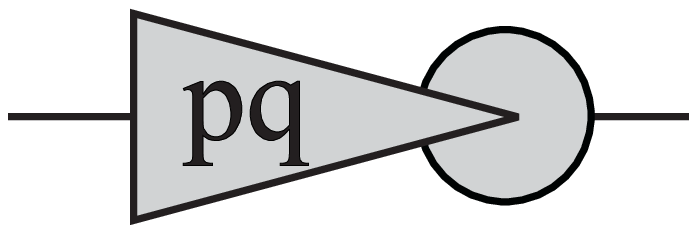}\equiv
\tilde{g}^{pq}(\omega)\tilde{B}^{pq}$ is the ``dressed'' Green's
functions with the renormalized vertex part.
We shall consider only the uniform phase in which all sites are equivalent.
Then the equation for the vertex part
$\tilde{B}^{pq}$ can be written as
\begin{equation}
\epsfxsize=14cm \epsfbox{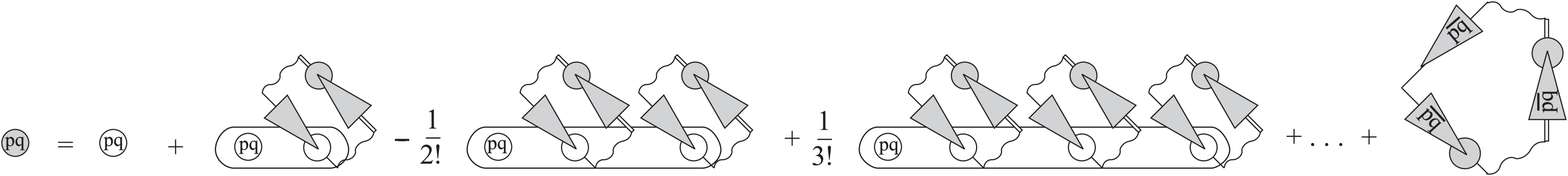}
\label{eq3}
\end{equation}
Via  $\overline{pq}$ we denoted the change of a spin projection in diagrams. Thus
$\overline{41}=31$, and vice versa, $\overline{31}=41$. ``Dressed'' Green's
functions with semiinvariants $\epsfxsize=2.5em \epsfbox{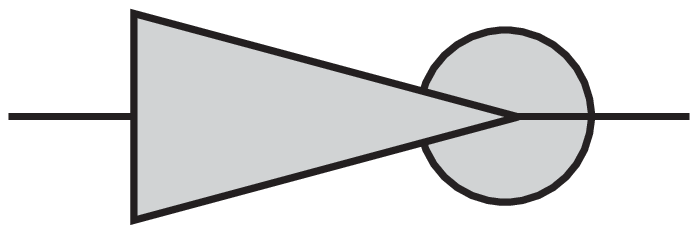}$ are of
$\tilde{g}^{41}(\omega)\tilde{B}^{41}$ or $\tilde{g}^{31}(\omega)\tilde{B}^{31}$ type.
They creates loops with spin $\uparrow$ or $\downarrow$, respectively.
The series in the right hand side of the equation is analogous to
the series which arises in the  mean field theory for the Ising or
Heisenberg model.

Formally, expression (\ref{eq3}) can be presented as
\begin{eqnarray}
\tilde{B}^{pq}&=&\frac{1}{Z}Sp\left(\hat{B}_{i}^{pq}{\rm e}^{-\beta
H_{eff}^{i}}\right) -\tilde{L}^{\overline{pq}},
\label{eq3a}
\end{eqnarray}
where
\begin{eqnarray}
H_{eff}^{i}&=&H_{0}+\tilde{Q}^{41} \hat{B}_{i}^{41} +
\tilde{Q}^{31}\hat{B}_{i}^{31}
\equiv H_{0}+\sum_{\{pq\}} \tilde{Q}^{pq}\hat{B}_{i}^{pq}, \;\; \{pq\}=41, 31;
\nonumber \\
Z&=&Sp\left({\rm e}^{-\beta H_{eff}^{i}}\right), \quad
\tilde{Q}^{pq}\equiv \epsfxsize=1.2cm \epsfbox{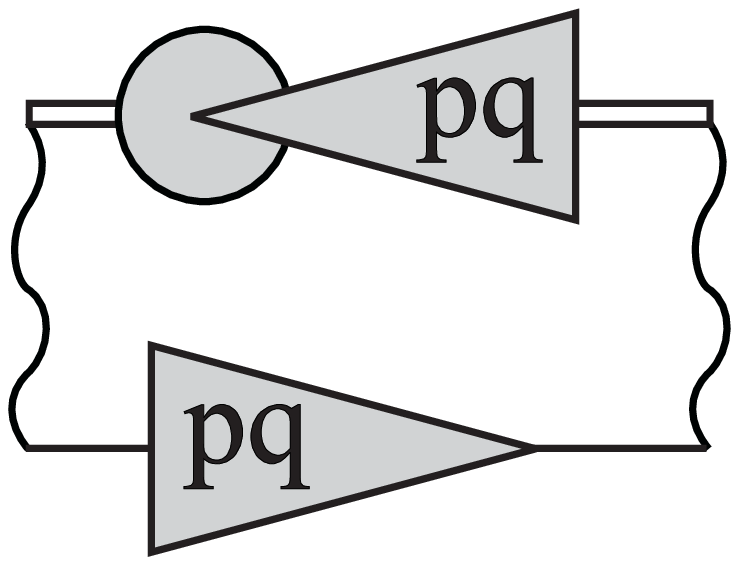}, \quad
\tilde{L}^{pq}\equiv \epsfxsize=1.2cm \epsfbox{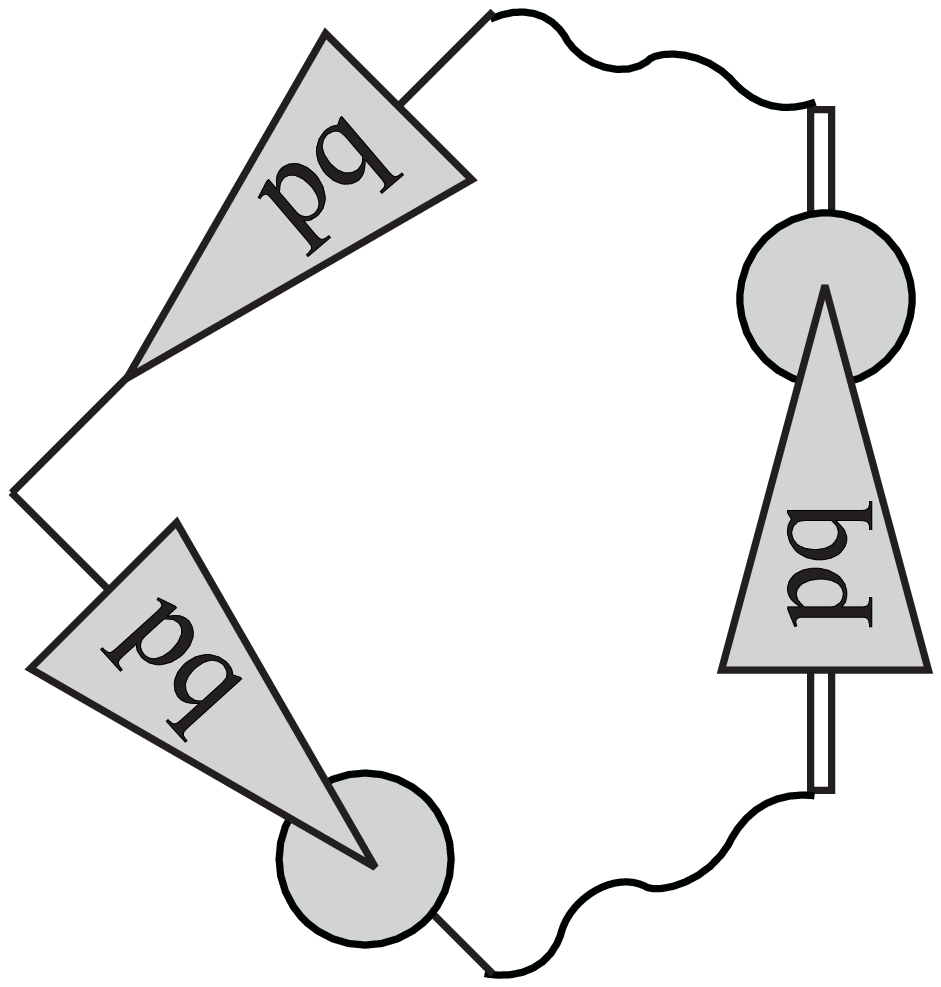}. 
\end{eqnarray}

The renormalized Green's functions $\tilde{g}^{pq}(\omega)$ is defined by the equation
\begin{eqnarray}
&&\epsfxsize=6 cm \epsfbox{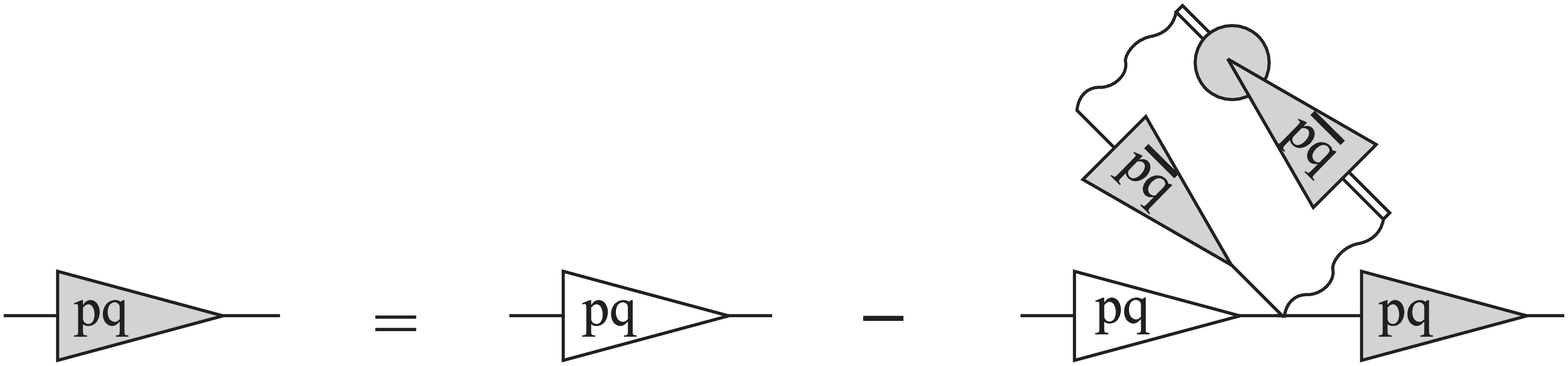} \\
&&{\rm Hence} \nonumber \\
&&\tilde{g}^{pq}=\frac{1}{{
i}\omega-\varepsilon^{pq}+\tilde{Q}^{\overline{pq}}}\equiv \frac{1}{{
i}\omega-\tilde{\varepsilon}^{pq}}, \quad \tilde{\varepsilon}^{pq}=
\varepsilon^{p}-\varepsilon^{q}-\tilde{Q}^{\overline{pq}}
\label{eq6}
\end{eqnarray}
That corresponds to the summation of diagrams of the 11, 12, 13 type.
The expression (\ref{eq6}) describes a renormalization of the single-electron
excitation spectrum.
$\tilde{Q}^{pq}$ plays the role of the irreducible self-energy
part, by which the levels of the single-electron transitions are shifted.
Using the renormalized Green's functions one can construct the loops
$\tilde{Q}^{pq}$ and $\tilde{L}^{pq}$ and,  thereby, derive an expression
for the irreducible part and equation for the auxiliary quantity
$\tilde{L}^{pq}$:
\begin{eqnarray}
\label{eq4}
\tilde{Q}^{pq}&=&
%-\frac{1}{2N} \sum_{k}t_{k}{\rm{th}}\frac{\beta}{2}
%(\tilde{\varepsilon}^{pq}+t_{k}\tilde{B}^{pq}) =
\frac{1}{N} \sum_{k}t_{k}{{n}}_+
(\tilde{\varepsilon}^{pq}+t_{k}\tilde{B}^{pq})
\qquad
\left (n_+(\varepsilon) = \frac{1}{1+{\rm e}^{\beta\varepsilon}} \right )
,
\\
\label{eq5}
%\end{eqnarray}
%\begin{eqnarray}
\tilde{L}^{pq}&=&
%\frac{1}{2}{\rm{th}}\frac{\beta}{2}\tilde{\varepsilon}^{pq}
%-\frac{1}{2N}\sum_{k}{\rm{th}}\frac{\beta}{2}
%(\tilde{\varepsilon}^{pq}+t_{k}\tilde{B}^{pq})=
%
\frac{1}{N} \sum_{k}{{n}}_+
(\tilde{\varepsilon}^{pq}+t_{k}\tilde{B}^{pq})
-
{{n}}_+
(\tilde{\varepsilon}^{pq})
.
\end{eqnarray}

The expression for the free energy was obtained with making use of
the analogy between the diagrammatic series for $\tilde{B}^{pq}$ and series
for the magnetization in the mean field theory for spin systems.
Then so called one-tail diagrams correspond to simple loops in current approximation.
One can
show that the free energy can be written as
\begin{eqnarray}
F&=&\mu n-T\ln Sp\left({\rm e}^{-\beta H_{eff}}\right)- \sum_{\{pq\}}
\tilde{Q}^{pq}\tilde{B}^{pq}+ \sum_{\{pq\}}\tilde{D}^{pq}, \nonumber\\
\tilde{D}^{pq}&=& \frac{1}{2}
\begin{array}{l}
\epsfxsize=1.2cm \epsfbox{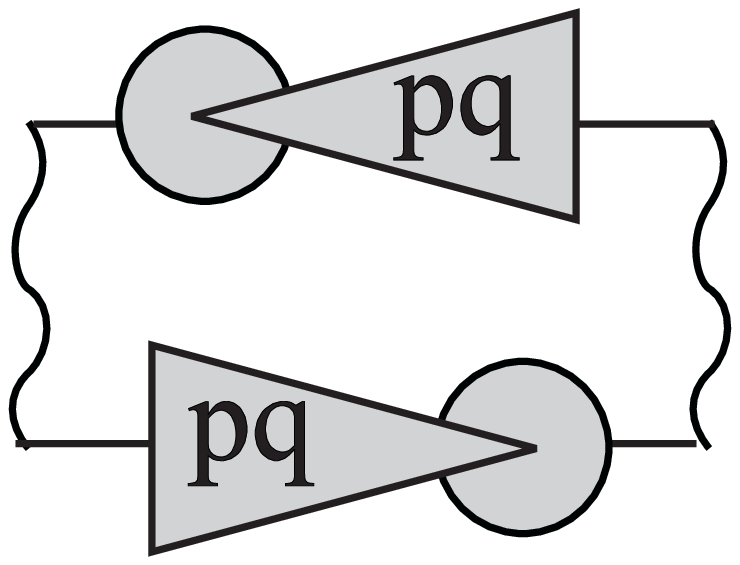}
\end{array}
+\frac{1}{3}
\begin{array}{l}
\epsfxsize=1.7cm \epsfbox{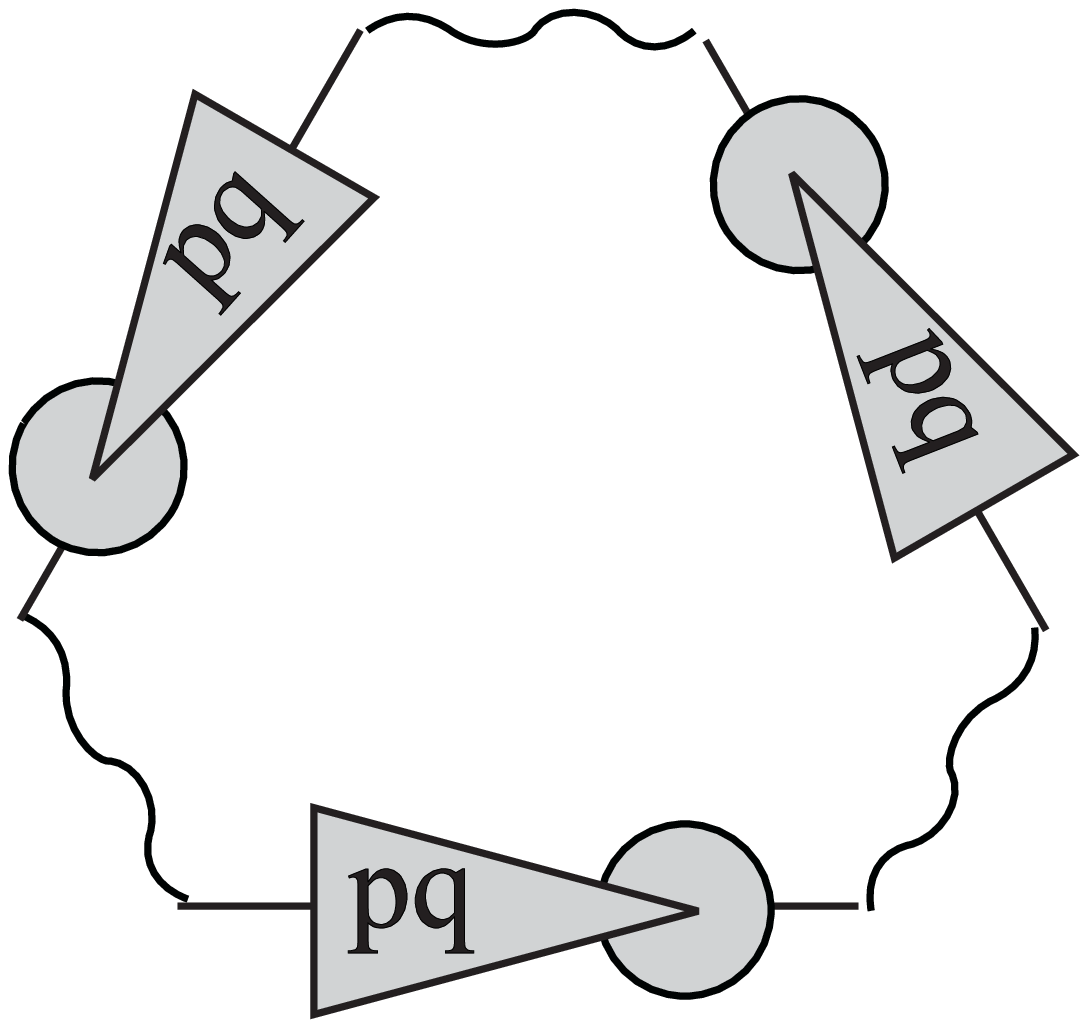}
\end{array}
+\frac{1}{4}
\begin{array}{l}
\epsfxsize=2.1cm \epsfbox{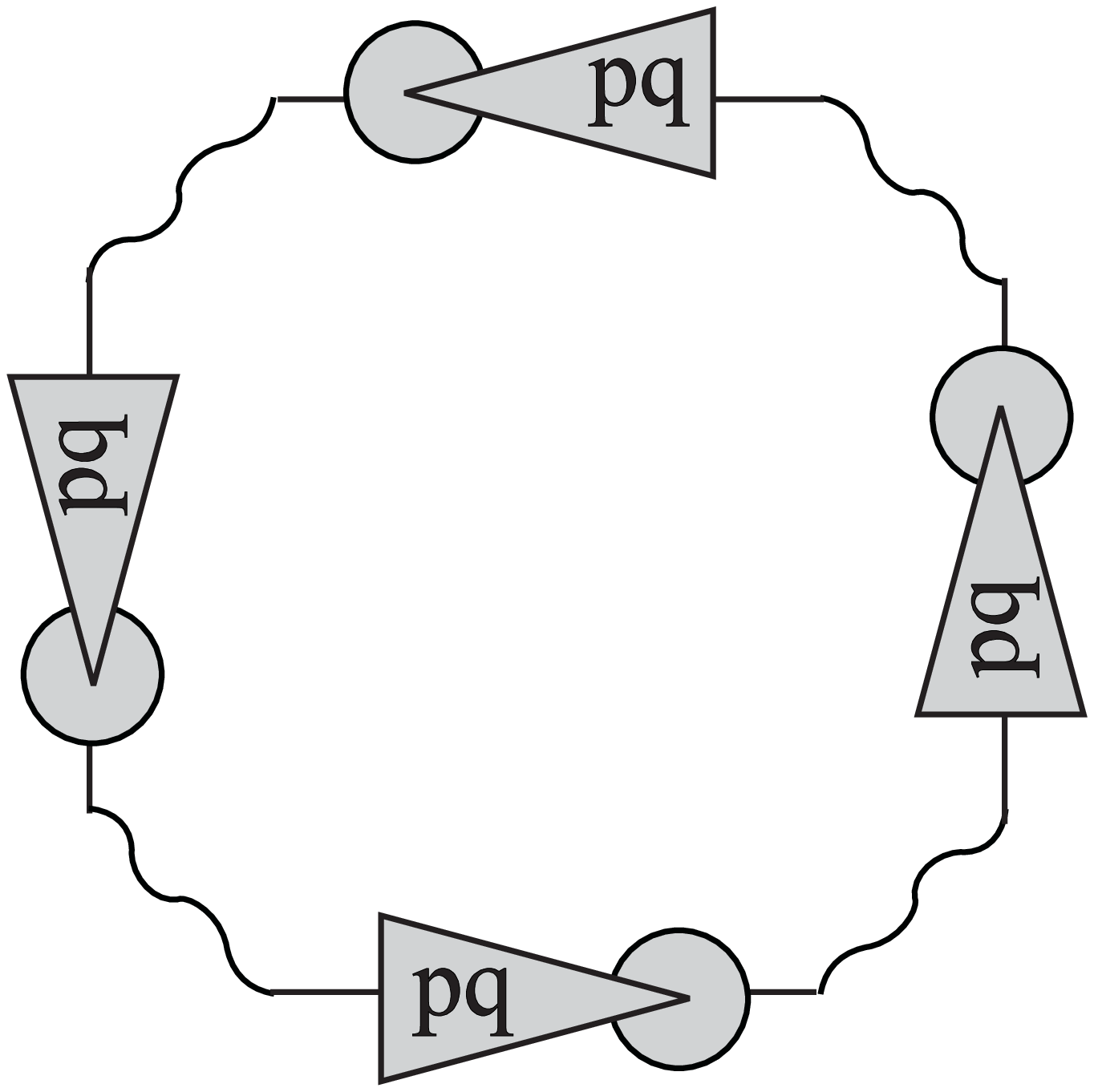}
\end{array}
+ \ldots
=\nonumber \\
&=&
%\frac{1}{\beta
%N}\sum_{k}\ln\frac{{\rm{ch}}\frac{\beta}{2}\tilde{\varepsilon}^{pq}}
%{{\rm{ch}}\frac{\beta}{2}(\tilde{\varepsilon}^{pq}+t_{k}\tilde{B}^{pq})}
%=
\frac{1}{\beta N} \sum_{k}\ln
\left( 1+{\rm e}^{-\beta \tilde{\varepsilon}^{pq}} \right)
-\frac{1}{\beta N} \sum_{k}\ln
\left( 1+{\rm e}^{-\beta
(\tilde{\varepsilon}^{pq}+t_{k}\tilde{B}^{pq})}\right).%
\label{eq7}
\end{eqnarray}
{\bf One can test this result by introducing additional fields
$h_{pq}\hat{B}^{pq}$ into Hamiltonian (\ref{eq1}). Then condition
$\frac{\partial F}{\partial {h}_{pq}}_{|h_{pq}=0}=\tilde{B}^{pq}$ will be
fulfilled, if one takes into account
(\ref{eq4}) and (\ref{eq5})}.

In the paramagnetic phase
 $\tilde{B}^{41}=\tilde{B}^{31}\equiv\tilde{B}$,
$\tilde{Q}^{41}=\tilde{Q}^{31}\equiv\tilde{Q}$. Hence,
the full set of simultaneous equations defining the mean field approximation
for thermodynamic quantities of Hubbard model consists of the equations (\ref{eq3a})
and (\ref{eq4}) as well as of {\bf the condition for mean number of electrons per
site
\begin{equation}
\langle X^{33}+X^{44}\rangle =n.
\end{equation}
One can show that standard condition for Hubbard operators
\begin{equation}
\langle X^{11}+X^{33}+X^{44}\rangle=\langle X^{44}\rangle +\tilde{B}=1
\end{equation}
is fulfilled automatically.

Considered this result within the wide class of approximation for Hubbard
model we noted that in the limit $U\rightarrow\infty$ it is equivalent to
results obtained within the moment technique \cite{Harris,Eskes}
(see also \cite{Pothoff})
which describes the shift of Hubbard bands on value depended on spin $\sigma$
\begin{equation}
B_{\sigma} = \frac{1}{n_{\sigma}(1-n_{\sigma})}\sum_{i\ne j}t_{ij}
\langle a_{i\sigma}^{+}a_{j\sigma}(2n_{i-\sigma}-1)\rangle.
\end{equation}
with $n_{\sigma}=\langle \hat{n}_{\sigma}\rangle$.
One can show that there is clear correspondence between $B_{\sigma}$ and
$\tilde{Q}^{pq}$:
\[
\tilde{Q}^{41} = -n_{\uparrow}B_{\uparrow}, \qquad
\tilde{Q}^{31} = -n_{\downarrow}B_{\downarrow}.
\]
But in MFA
approach we have also equations for end parts for Green functions
$\tilde{B}^{pq}$ ($n_{\sigma}$) which is needed for consideration of
processes where ordered phase takes place as well as the equation for
free energy (\ref{eq7}) required for investigation of phase transitions of
first order.}

\section{Two-sublattice Hubbard model with local anharmonicity}

The Hamiltonian of the two-sublattice
Hubbard model with local anharmonicity which includes also
an interaction of electrons with local anharmonic vibrations,
described within the pseudospin formalism, reads (\cite{DanStas98})
\begin{eqnarray}
&&H=H_{e}+H_{s}+H_{e-s}+H_{s-s}\nonumber\\
&&H_{e}=-\mu\sum_{i,s}(\hat{n}_{i1}^{s}+\hat{n}_{i2}^{s})+U \sum_{i}
(\hat{n}_{i1}^{\uparrow}\hat{n}_{i1}^{\downarrow}+ \hat{n}_{i2}^{\uparrow}
\hat{n}_{i2}^{\downarrow})+\sum_{ij}\sum_{s\alpha} t_{ij}
a_{is\alpha}^{+}a_{js\alpha}\nonumber\\
&&H_{s}=-h\sum_{i}(S_{i1}^{z}-S_{i2}^{z})\nonumber\\
&&H_{e-s}=g\sum_{is}(\hat{n}_{i1}^{s}S_{i1}^{z}-\hat{n}_{i2}^{s}S_{i2}^{z})
\nonumber\\
&&H_{s-s}=-\frac{1}{2} \sum_{i,j}\sum_{\alpha\beta} J_{ij}^{\alpha\beta}
S_{i\alpha}^{z}S_{j\beta}^{z}
\label{eq14}
\end{eqnarray}
Here $H_{e}$ is the Hubbard Hamiltonian for two sublattices
(subplanes in the case of the YBa$_2$Cu$_3$O$_{7-\delta}$ structure),
$H_{s}+H_{s-s}$
is a pseudospin part of the Hamiltonian similar to the Mitsui
model used in the ferroelectrics theory; $H_{e-s}$ describes the
interaction between the electron and pseudospin subsystems.
The operators $\hat{n}_{i\alpha}^{s}$ and $S_{i\alpha}^{z}$
act at the site $i$
of the plane $\alpha$ ($\alpha$=1,2). $h$ describes the asymmetry of a
two-minima potentials.

Similarly to that performed for the standard Hubbard model (\ref{eq1}), let
us introduce a cluster basis of states $\mid R_{i}\rangle\equiv \mid
n_{i1}^{\uparrow}, n_{i1}^{\downarrow}, n_{i2}^{\uparrow},
n_{i2}^{\downarrow}\rangle$ $\otimes$ $\mid S_{i1}^{z}, S_{i2}^{z}\rangle$,
 or $\mid R_{i}\rangle$ $\equiv$ $\mid n_{i1}^{\uparrow},
n_{i1}^{\downarrow},n_{i2}^{\uparrow}, n_{i2}^{\downarrow},
S_{i1}^{z},S_{i2}^{z}\rangle$ (here $n_{i\alpha}^{\uparrow},
n_{i\alpha}^{\downarrow}$ denote eigenvalues of the operators
$\hat{n}_{i\alpha}^{\uparrow}$ and$\hat{n}_{i\alpha}^{\downarrow}$),
consisting of 64 vectors of state
\begin{equation}
\begin{array}{l}
\mid 1\rangle =\mid 0,0,0,0,\uparrow,\uparrow\rangle \\
\ldots \\
\mid 64\rangle = \mid 1,1,1,1,\downarrow,\downarrow\rangle.
\end{array}
\end{equation}
Analysis of the thermodynamics of the model
(\ref{eq14}) without taking into account the electron transfer has been
performed in \cite{DanStas98}, where the regions of ferroelectric pseudospin
ordering were found and phase diagrams where built. Now we will investigate the influence
of electron transfer on thermodynamic properties and phase transitions.
Let us consider a pseudospin part of the
Hamiltonian within the mean field approximation and an electron subsystem
via the presented in the previous Section selfconsistent GRPA scheme.

The cluster basis of states consists of a direct product of two
single-sublattice
{$\mid i,\alpha,R\rangle$} $\equiv$ $\mid \hat{n}_{i\alpha}^{\uparrow},
\hat{n}_{i\alpha}^{\downarrow}, \hat{S}_{i\alpha}^{z}\rangle$
($\alpha=1,2$) ones.
Each single-sublattice set of states includes eight components at a
given site
$i$ and sublattice $\alpha$:
\begin{equation}
\begin{array}{ll}
\mid 1\rangle =\mid 0, 0,
\uparrow\rangle  &   \mid \tilde{1}\rangle =\mid 0,0,
\downarrow\rangle \\
\mid 2\rangle = \mid 1, 1,
\uparrow\rangle  &   \mid \tilde{2}\rangle = \mid
1,1,\downarrow\rangle \\
\mid 3\rangle = \mid 0,1,\uparrow\rangle  &   \mid\tilde{3}\rangle =
\mid 0,1,\downarrow\rangle \\
\mid 4\rangle = \mid 1,0,\uparrow\rangle  &   \mid\tilde{4}\rangle =
\mid 1,0,\downarrow\rangle
\end{array}
\label{eq15}
\end{equation}

Let us introduce the Hubbard operators $X_{i\alpha}^{RS}=\mid
i,\alpha,R\rangle\langle i,\alpha,S\mid$ in this basis. Then the
operators $a_{i\uparrow\alpha}$, $a_{i\downarrow\alpha}$, $S_{i\alpha}^{z}$
are related  to the X-operators as
\begin{eqnarray}
&&a_{i\uparrow\alpha}=X_{i\alpha}^{14}+X_{i\alpha}^{32}+
X_{i\alpha}^{\tilde{1}\tilde{4}}+X_{i\alpha}^{\tilde{3}\tilde{2}}, \quad
a_{i\downarrow\alpha}= X_{i\alpha}^{13}-X_{i\alpha}^{42} +
X_{i\alpha}^{\tilde{1}\tilde{3}} - X_{i\alpha}^{\tilde{4}\tilde{2}}
\nonumber\\
&&S_{i\alpha}^{z}=\frac{1}{2}\sum_{k=1}^{4}(X_{i\alpha}^{RR}-
X_{i\alpha}^{\tilde{R}\tilde{R}})
\label{eq16}
\end{eqnarray}
Then Hamiltonian (\ref{eq14}) can be easily rewritten as
\begin{eqnarray}
&&H=H_{0}+ H_{int}+H_{s-s},
\end{eqnarray}
where $H_{0}=\sum\limits_{p=1}^{\tilde{4}}\sum\limits_{i,\alpha}
\varepsilon_{\alpha}^{p} X_{i\alpha}^{pp}$ with
\begin{eqnarray}
&&
\begin{array}{ll}
\varepsilon_{1}^{1,\tilde{1}}=\mp\frac{h}{2}, \quad
&\varepsilon_{2}^{1,\tilde{1}}=\pm\frac{h}{2} \\
\varepsilon_{1}^{2,\tilde{2}}=-2\mu +U\mp\frac{h}{2}\pm g, \quad
&\varepsilon_{2}^{2,\tilde{2}}=-2\mu +U\pm\frac{h}{2}\mp g \\
\varepsilon_{1}^{3,\tilde{3}}=\varepsilon_{1}^{4,\tilde{4}} =-\mu\mp
\frac{h}{2} \pm\frac{g}{2},
&\varepsilon_{2}^{3,\tilde{3}}=
\varepsilon_{2}^{4,\tilde{4}}=-\mu\pm\frac{h}{2}\mp\frac{g}{2}
\end{array}
\end{eqnarray}
and
\begin{eqnarray}
&&H_{int} =\sum_{ij,\alpha} t_{ij} (X_{i\alpha}^{41} + X_{i\alpha}^{23} +
X_{i\alpha}^{\tilde{4}\tilde{1}} + X_{i\alpha}^{\tilde{2}\tilde{3}})\times
(X_{j\alpha}^{14} + X_{j\alpha}^{32} +X_{j\alpha}^{\tilde{1}\tilde{4}} +
X_{j\alpha}^{\tilde{2}\tilde{3}}) + \nonumber\\
&& + \sum_{ij\alpha} t_{ij} (X_{i\alpha}^{31} - X_{i\alpha}^{24}
+X_{i\alpha}^{\tilde{3}\tilde{1}} - X_{i\alpha}^{\tilde{2}\tilde{4}})\times
(X_{j\alpha}^{13} - X_{j\alpha}^{42} + X_{j\alpha}^{\tilde{1}\tilde{3}} -
X_{j\alpha}^{\tilde{4}\tilde{2}}),
\label{eq17}
\end{eqnarray}
$H_{s-s}$ is left unchanged.

It should be noted that  the model (\ref{eq14}) does not take into account
the transfer of electrons between the planes. Therefore, if the
interplane interaction between pseudospins is absent
$(J_{ii'}^{12}=J_{ii'}^{21}=0)$, and also owing to the system
symmetry, the problem is reduced to two identical single-sublattice
subproblems.

Due to the relation
\begin{equation}
[S_{i\alpha}^{z}, X_{j\beta}^{pq}]=0,
\end{equation}
and also when a splitting of the mean field type
$S_{i\alpha}^{z}S_{j\beta}^{z}=\langle S^{z}_{i\alpha}\rangle S_{j\beta}^{z}
+\langle S_{j\beta}^{z}\rangle S_{i\alpha}^{z}-\frac{1}{2} \langle
S_{i\alpha}^{z}\rangle \langle S_{j\beta}^{z}\rangle$ is used, the
effective Hamiltonian of the problem  reads
\begin{equation}
H_{eff}=H_{0} +\sum_{\{pq\}}\sum_{i\alpha}
\tilde{Q}_{\alpha}^{pq} \hat{B}_{i\alpha}^{pq} -
\sum_{i\alpha\alpha'} J_{\alpha\alpha'} \langle S_{\alpha'}^{z}\rangle
S_{i\alpha}^{z},
\label{eq18}
\end{equation}
where $J_{\alpha\alpha'}=\sum\limits_{i}J_{ii'}^{\alpha\alpha'}$.

When writing equations of self-consistent GRPA approach for the two-sublattice model
 (\ref{eq14}) in the approximation of independent subbands \cite{SS94} and
 at $U=\infty$, we can use expressions of the previous Section. All
 the quantities $\tilde{Q}_{\alpha}^{pq}$, $\tilde{B}_{\alpha}^{pq}$, etc.
get an auxiliary index indicating the subplane $\alpha=1,2$. For instance,
owing to the presence of pseudospin degrees of freedom, in addition to
$B_{\alpha}^{pq}$ we also have $B_{\alpha}^{\tilde{p}\tilde{q}}$. Then the
summation $\sum\limits_{\{pq\}}$ means the summation over $pq = 41,
\tilde{4}\tilde{1}, 31, \tilde{3}\tilde{1}$. An additional contribution to
the free energy $\tilde{D}^{pq}$ we should change by
$\tilde{D}^{pq}_{\alpha}$, where
\begin{equation}
\tilde{D}_{\alpha}^{pq}=
%\frac{1}{\beta N}\sum\limits_{n}\ln
%\frac{{\rm{ch}}\frac{\beta}{2}\tilde{\varepsilon}_{\alpha}^{pq}}
%{{\rm{ch}}\frac{\beta}{2}(\tilde{\varepsilon}_{\alpha}^{pq}+
%\tilde{B}_{\alpha}^{pq}t_{n})}=
\frac{1}{\beta N} \sum_{k}\ln
\left( 1+{\rm e}^{-\beta \tilde{\varepsilon}_{\alpha}^{pq}} \right)
-\frac{1}{\beta N} \sum_{k}\ln
\left( 1+{\rm e}^{-\beta
(\tilde{\varepsilon}_{\alpha}^{pq}+t_{k}\tilde{B}_{\alpha}^{pq})}\right).
,
\end{equation}
with
\[
\tilde{\varepsilon}_{\alpha}^{pq}=\varepsilon_{\alpha}^{p}-
\varepsilon_{\alpha}^{q}- \tilde{Q}_{\alpha}^{\overline{pq}}.
\]

The expression for a free energy now reads
\begin{equation}
F=\mu n -T\ln Sp \left({\rm e}^{-\beta(H_{eff})}\right)
-\sum\limits_{{\{pq\}}\atop{\alpha}}
\tilde{Q}_{\alpha}^{pq} \tilde{B}_{\alpha}^{pq} +
\sum\limits_{{\{pq\}}\atop{\alpha}}
\tilde{D}_{\alpha}^{pq} +\frac{1}{2}\sum_{\alpha\alpha'}
J_{\alpha\alpha'} \langle S_{\alpha}^{z}\rangle \langle
S_{\alpha'}^{z}\rangle.
\end{equation}
The system of equations to be solved
\begin{equation}
\begin{array}{l}
\tilde{B}_{\alpha}^{pq}=\frac{Sp(\hat{B}_{\alpha}^{pq} {\rm e}^{-\beta
H_{eff}})} {Sp({\rm e}^{-\beta H_{eff}})}-\tilde{L}^{\overline{pq}} \nonumber\\

\tilde{Q}_{\alpha}^{pq}=
%-\frac{1}{2N} \sum_{n}t_{n}{\rm{th}} \frac{\beta}{2}
%(\tilde{\varepsilon}_{\alpha}^{pq} + t_{n}\tilde{B}_{\alpha}^{pq})
\frac{1}{N} \sum\limits_{k}t_{k}{{n}}_+
(\tilde{\varepsilon}_{\alpha}^{pq}+t_{k}\tilde{B}_{\alpha}^{pq})

\end{array}
\end{equation}
should be supplemented by a standard identity for Hubbard operators
$\langle \sum\limits_{p=1}^{\tilde{4}} X_{\alpha}^{pp}
\rangle =1 $, by the equation for the pseudospin mean value
$\langle S_{\alpha}^{z}\rangle= Sp(\hat{S}^{z}_{\alpha}{\rm e}^{-\beta
H_{eff}})/Sp({\rm e}^{-\beta H_{eff}})$,
the equation for the chemical potential
$\sum\limits_{\alpha=1}^2\langle X_{\alpha}^{33} + X_{\alpha}^{44} +
X_{\alpha}^{\tilde{3}\tilde{3}} +
X_{\alpha}^{\tilde{4}\tilde{4}}\rangle_{H_{eff}}=n$
as well as by the equalities
$\langle X_{\alpha}^{33}\rangle = \langle X_{\alpha}^{44}\rangle$,
$\langle X_{\alpha}^{\tilde{3}\tilde{3}}\rangle = \langle
X_{\alpha}^{\tilde{4}\tilde{4}}\rangle$ in paramagnetic case.

\section{Numerical calculations and results}

In paper \cite{SH98}, where the pseudospin-electron model with zero transfer
was studied, the existence  of the phase separation resulting
from the direct interaction between pseudospins was revealed.
This effect takes place, as it was shown, in the certain region of the model
parameters values in the regime of the fixed electron concentration ($n=const$).
On the other hand, the simple pseudospin-electron model or the more complicated
model (\ref{eq14}) introduced
for a description of possible phase instabilities in HTSCs of the
YBaCuO type do not take into account explicitly the existence of
CuO chains which are
a reservoir supplying a charge carriers into the superconducting planes.

Therefore, the model can be considered in two regimes:
a) constant concentration, and b) constant chemical potential
(the fixed value of the chemical potential  $\mu$ is sustained by structural
elements which are not taken into account within the model).

\begin{figure}
\begin{center}
\begin{minipage}{8cm}
\epsfxsize=8cm
\epsfbox{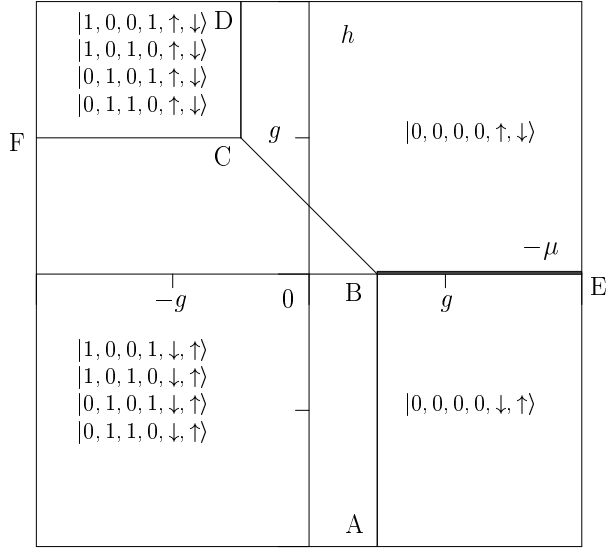}
\end{minipage}
\end{center}
\caption{Ground state diagram in the
$\mu$-$h$ plane in absence of direct pseudospin ($J_{\alpha\beta}=0$) interaction
($J_{\alpha\beta}=0$) and electron transfer ($t_{ij}=0$). $U=\infty$.
The electron and pseudospin configurations are shown.}
\label{fig:0}
\end{figure}

The ground state diagram in absence of direct pseudospin-pseudospin interaction and transfer
($J_{\alpha\beta}=0, t_{ij}=0$) is shown in Fig.\ref{fig:0}. At
$T=0$  the chemical potential coincide with the line ABCD at any value
of concentration $n$. When temperature is different from zero, the
chemical potential lies to the right of ABCD at $n\sim 0$ and to
the left of it at
$n\sim 2$. In the $\mu = const$ regime, the chemical potential is given by the
line parallel to the ordinate axis.

Analysis performed in the spirit of the mean field theory \cite{DanStas98} has shown that
in the regime of a fixed value of the chemical potential, the region of
ferroelectric type ordering of pseudospins lies in the vicinity of the FCBE curve. This curve is
a boundary between the ground states with the pseudospin configurations
$|\uparrow\downarrow\rangle$ and
$|\downarrow\uparrow\rangle$. If the interplane interaction  $J_{12}$ is
different from zero, the mean value
$\eta=\langle S_1^z+S_2^z\rangle$ is the order parameter: $\eta=0$ in the
paraphase
 and $\eta\ne 0$ in the ferroelectric phase. At $J_{12}=0$ we deal with two
 independent single-sublattice problems, and the order parameter is zero;
however, a jump of the parameter $\xi=\langle S_1^z-S_2^z\rangle$ takes
place.

\begin{figure}
\begin{center}
\begin{minipage}{7cm}
\epsfxsize=7cm \epsfbox{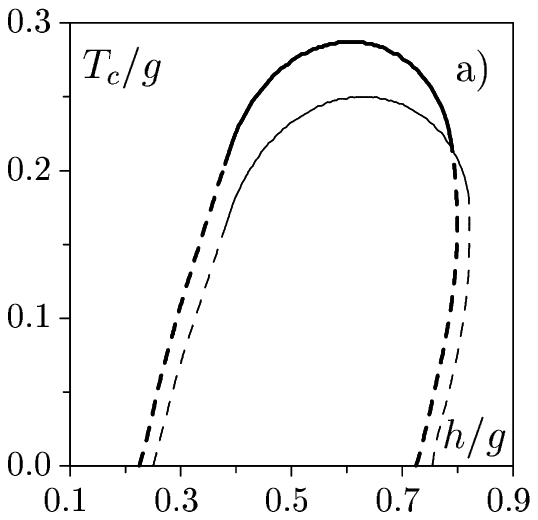}
\end{minipage}\hspace{1em}\begin{minipage}{7cm}
\epsfxsize=7cm \epsfbox{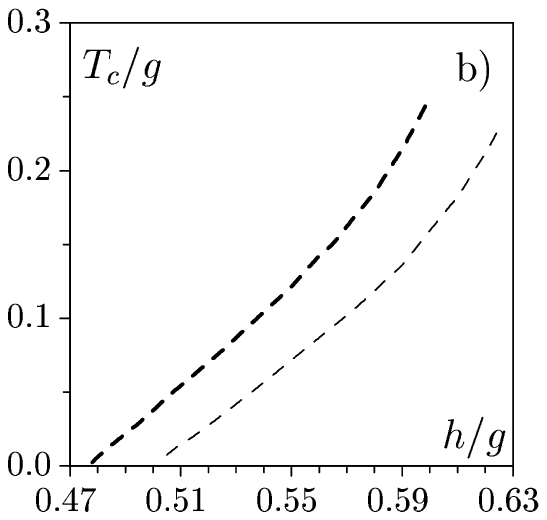}
\end{minipage}
\end{center}
\caption{Dependence of temperature of the phase
transition
$T_c$ on the parameter $h$ at different values of interaction parameters
$J_{11}$ and $J_{12}$ in the regime
$\mu=const=0$.
Thick lines correspond to the case $t_{ij}/g=0.2$; thin lines
correspond to the case $t_{ij}=0$.
a) $J_{11}=J_{12}=g/2$, b) $J_{11}=g, J_{12}=0$.  Phase transitions are
 of the second order (solid lines) or of the first order (dashed lines).}
\label{fig:2}
\end{figure}

%
%---------------------
\begin{figure}
\begin{center}
\begin{minipage}{7cm}
\epsfxsize=7cm \epsfbox{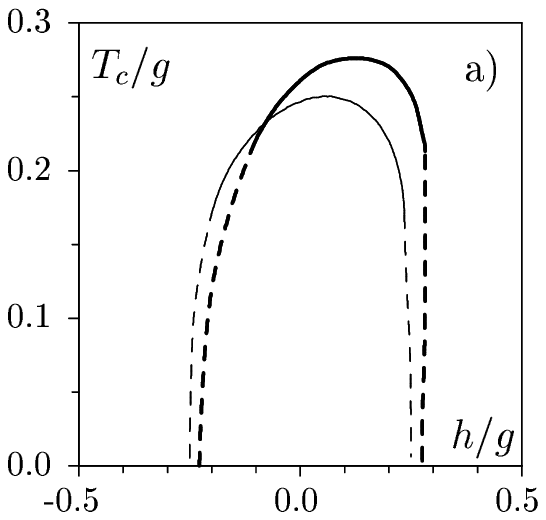}
\end{minipage}\hspace{1em}\begin{minipage}{8cm}
\epsfxsize=8cm \epsfbox{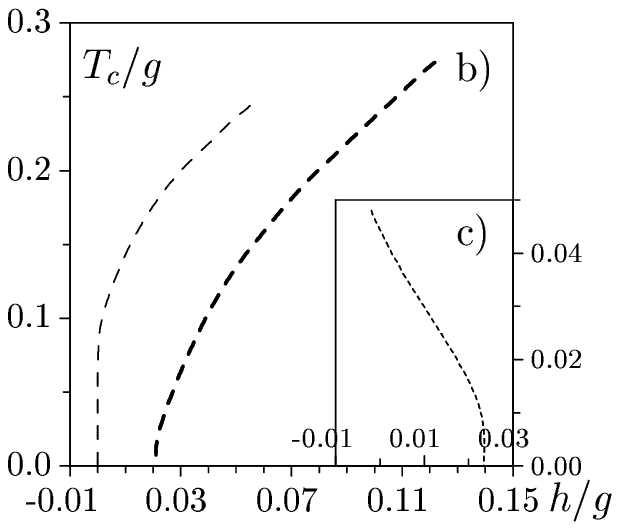}
\end{minipage}
\end{center}
\caption{Dependence of temperature of the phase
transition
$T_c$ on the parameter $h$ at different values of
interaction parameters $J_{11}$ and $J_{12}$ in the regime
 $\mu=const=-g$.
Thick lines correspond to the case  $t_{ij}/g=0.2$; thin lines
correspond to the case
$t_{ij}=0$.
a) $J_{11}=J_{12}=g/2$,
b) $J_{11}=g, J_{12}=0$.
c) $J_{11}=J_{12}=0, t_{ij}/g=0.2$. Phase transitions are
 of the second order (solid lines) or of the first order (dashed lines).}

\label{fig:3}
\end{figure}

Figs.\ref{fig:2},\ref{fig:3} illustrate the influence of electron
subsystem on a appearance of the phase diagrams. Let us note
that in the Mitsui model (to which one can pass by putting $g=0$ and
$t_{ij}=0$) the phase diagram has a form of a curve symmetric with
respect to  $h$. At ``switching on'' the interaction between the
electron and pseudospin subsystems $g$, the center of a region with
ferroelectric ordering shifts from $0$ to $g$  in accordance with the
broken line FCBE position when the chemical potential is changed (Fig.\ref{fig:0}).
Besides, the phase diagram becomes asymmetric with respect to the center
of the pseudospin ordering region. As it has been already noted, at
$J_{12}=0$ the ferroelectric phase does not exist
(figs.\ref{fig:2}b,\ref{fig:3}b,\ref{fig:3}c),  a discontinuous change of
$\xi$ takes place.

Let us now consider the influence of electron transfer on the phase transitions
in the simplest approximation Hubbard-I. It has been shown in \cite{SS93}  that at
$U\rightarrow\infty$ in a one-sublattice case for the Hubbard model with
local anharmonicity, one has in Hubbard-I approximation two subbands, separated by the
gap equal to $g$:
$\varepsilon^{31}({\bf k})=\varepsilon^{41}({\bf k})=\frac{g}{2}+
t_{\bf k}\langle X^{44}+X^{11}\rangle_0$ and
$\varepsilon^{\tilde 3\tilde 1}({\bf k})=\varepsilon^{\tilde 4\tilde 1}
({\bf k})=-\frac{g}{2}+
t_{\bf k}\langle X^{\tilde 4\tilde 4}+X^{\tilde 1\tilde 1}\rangle_0$. In the
considered in this paper approximation, these subbands are shifted by
$\tilde{Q}^{pq}$; the mean values of the Hubbard operators are not
calculated in the zero-order approximation but found self-consistently.
In the case shown on Fig.\ref{fig:2} ($\mu=0$), the
chemical potential lies right in between of the mentioned bands. The
number of charge carriers slightly changes with temperature and parameter
$h$.  The transfer can facilitate, but by itself is not able to cause
the phase transition.

\begin{figure}

\begin{center}
\begin{minipage}{7cm}
\epsfxsize=7cm \epsfbox{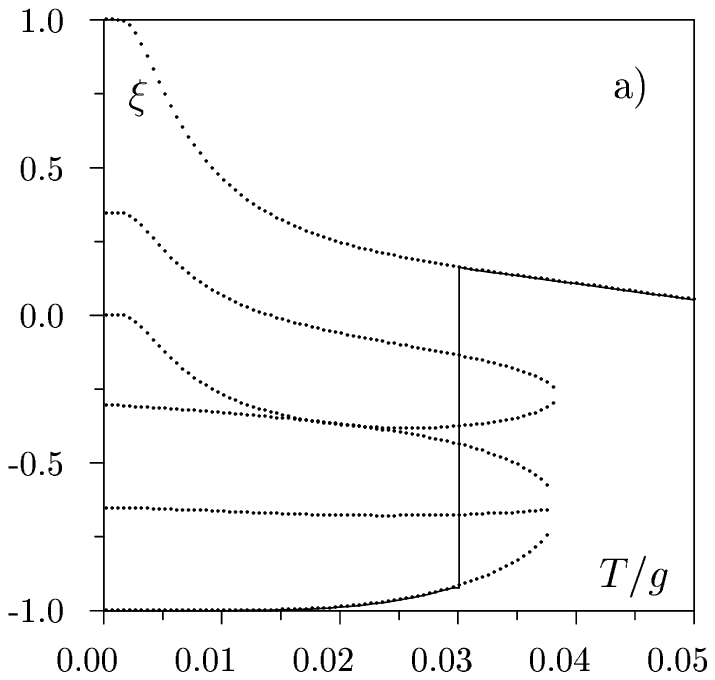}
\end{minipage}\hspace{1em}\begin{minipage}{7cm}
\epsfxsize=7cm \epsfbox{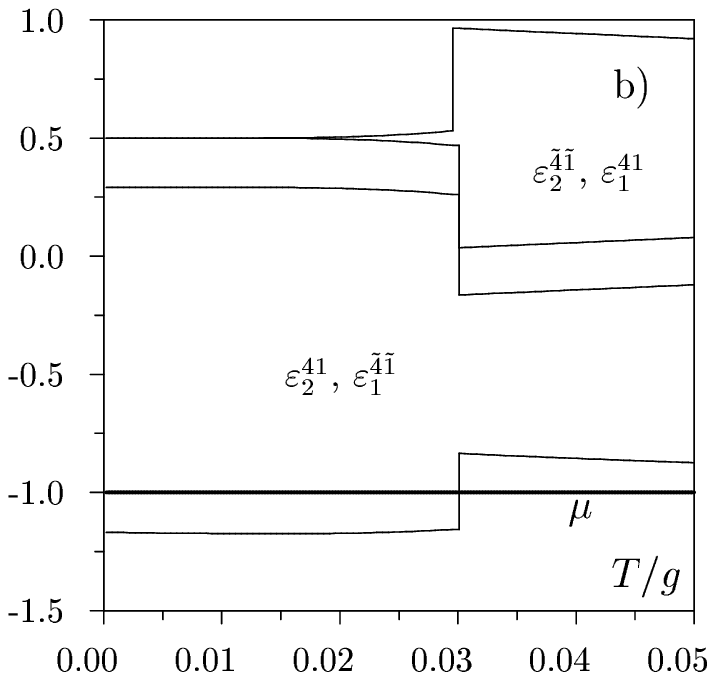}
\end{minipage}
\begin{minipage}{7cm}
\epsfxsize=7cm \epsfbox{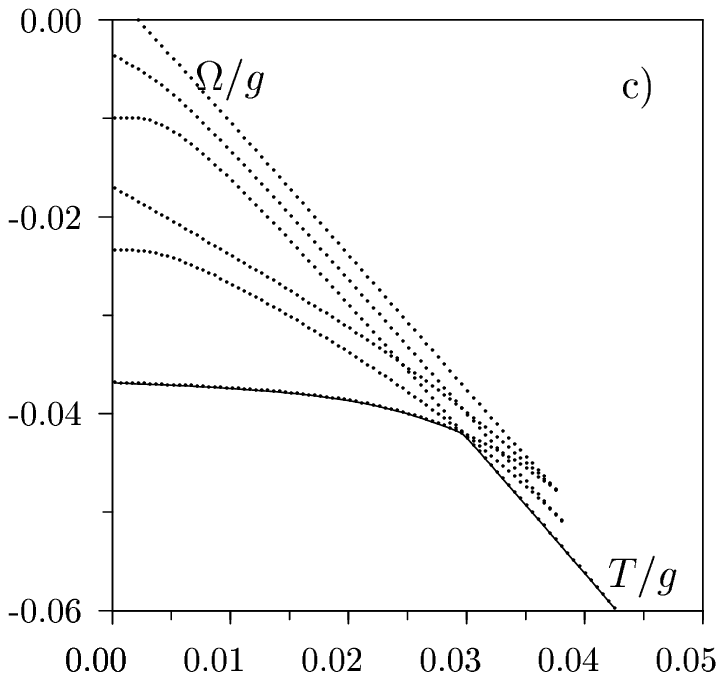}
\end{minipage}\hspace{1em}\begin{minipage}{7cm}
\epsfxsize=7cm \epsfbox{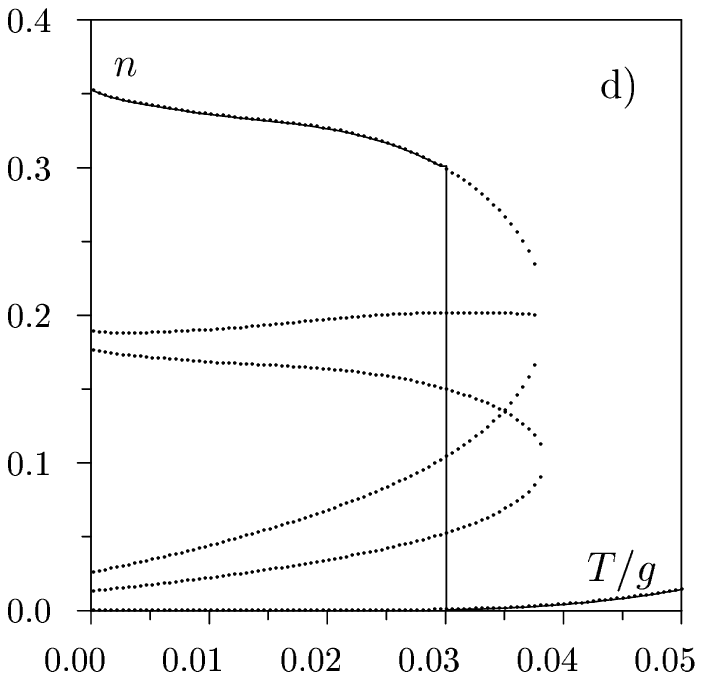}
\end{minipage}
\end{center}

\caption{
Dependence of the order parameter $\xi$ (a),
band edges (b),
thermodynamic potential (c), and electron concentration (d)
on temperature in the regime $\mu=const=-g$. The parameters values
are:
$J_{11}=J_{12}=0$, $T/g=0.1$, $t_{ij}/g=0.2$, $h/g=0.01$.
}
\label{fig:4}
\end{figure}

At $\mu=-g$ (Fig.\ref{fig:3}) the chemical potential lies near the
edge of the lower band. Even unsignificant change in the
spectrum can make the chemical potential leave the
band (Fig.\ref{fig:4}b) and provokes jump of parameter $\xi$
(Fig.\ref{fig:4}a). In this case,
interaction via electron transfer alone can cause
the phase transition in the system.
Fig.\ref{fig:4}c indicates that the phase transition of the first order takes place.
The behaviour of electron
concentration here is quite specific (Fig.\ref{fig:4}d), when it jumps from a certain value at
$T=0$ up to $0$ when the chemical potential leaves the band. On the
other hand, the existence of instable and metastable branches in
Fig.\ref{fig:4} indicates that the chemical potential is complicated
function of concentration. Therefore the situation when at a certain
value of $n$ the function $\mu(n)$ has a jump becomes possible. In this case
we have separation to two phases with different concentration, similarly
to  a seen in the $P(V)$ diagram separation of vapour into gas and
liquid in the gas-liquid phase transition.

\begin{figure}

\begin{center}
\begin{minipage}{7cm}
\epsfxsize=6.6cm \epsfbox{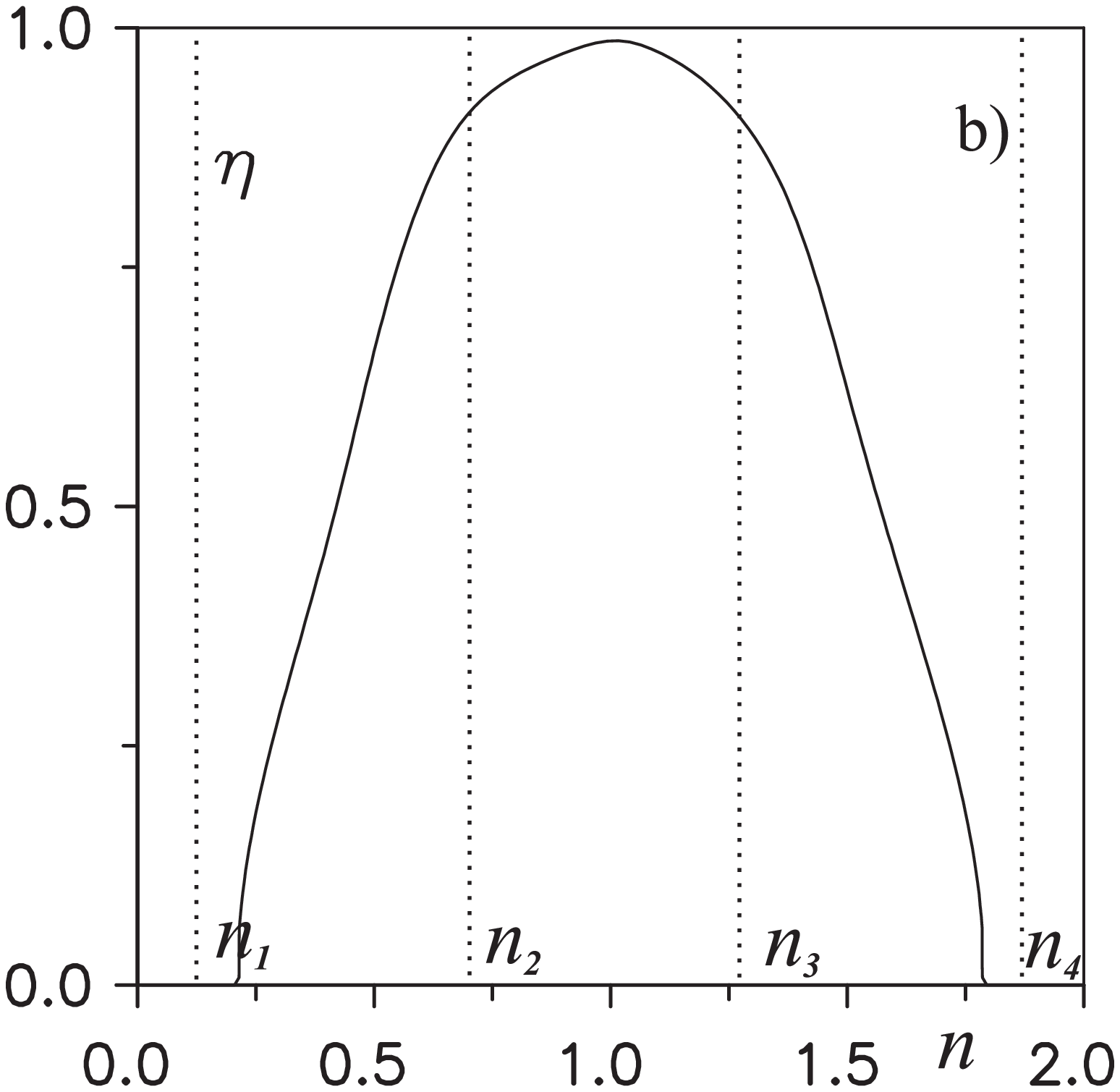}
\end{minipage}\hspace{1em}\begin{minipage}{8cm}
\epsfxsize=8cm \epsfbox{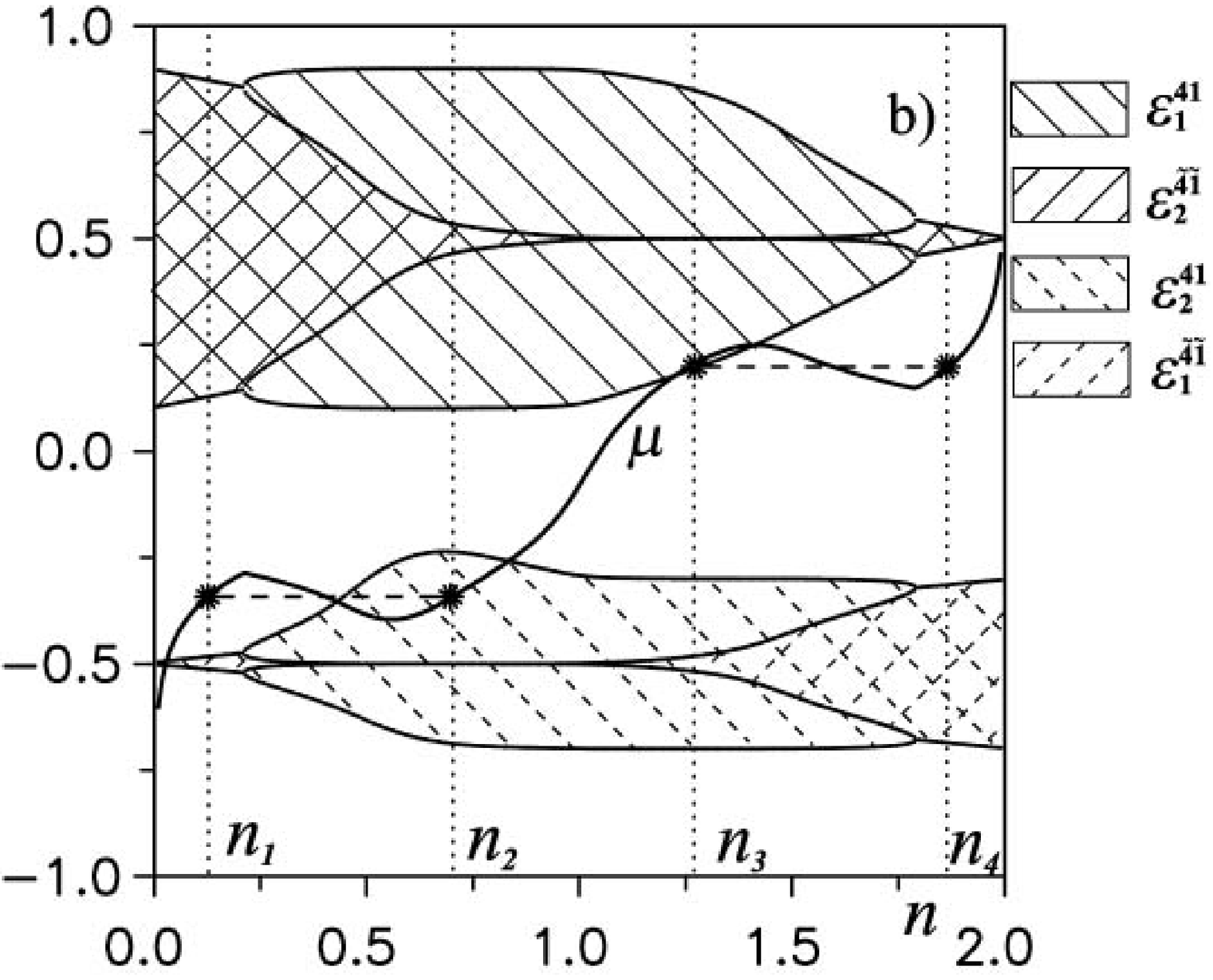}
\end{minipage}
\epsfxsize=7cm \epsfbox{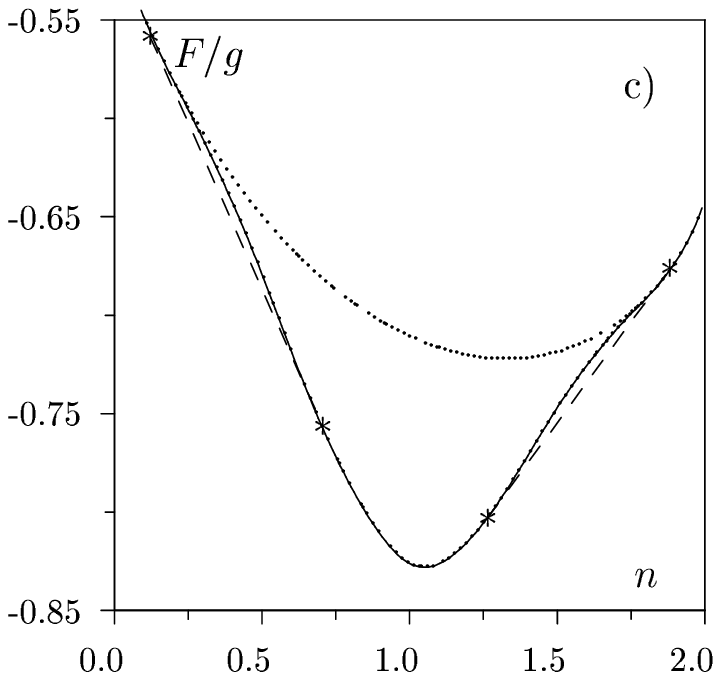}
\end{center}

\caption{
Dependence of the order parameter $\eta$ (a),
band spectrum (b), and free energy (c) on concentration. The parameters
values are:
$J_{11}=J_{12}=g/2$, $T/g=0.1$, $t_{ij}/g=0.1$, $h/g=0.5$.
}
\label{fig:5}
\end{figure}

\begin{figure}

\begin{center}
\begin{minipage}{7cm}
\epsfxsize=7cm \epsfbox{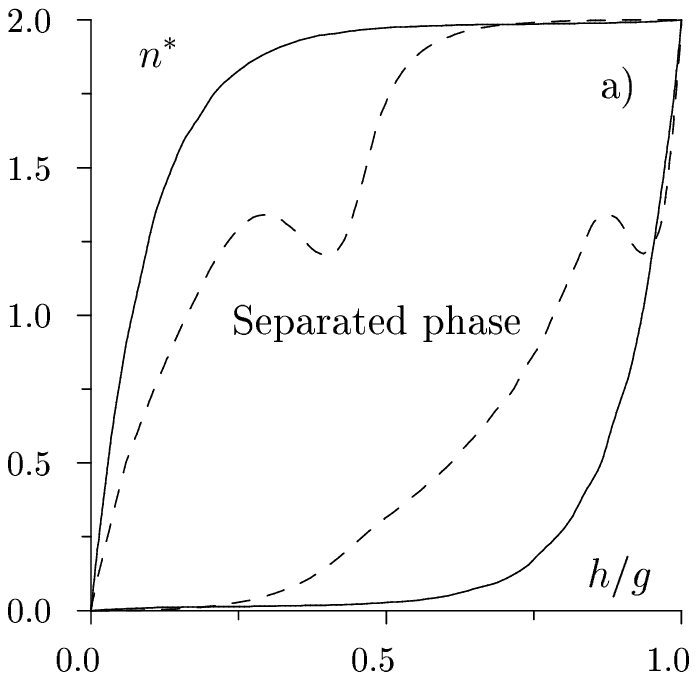}
\end{minipage}\hspace{1em}\begin{minipage}{7cm}
\epsfxsize=7cm \epsfbox{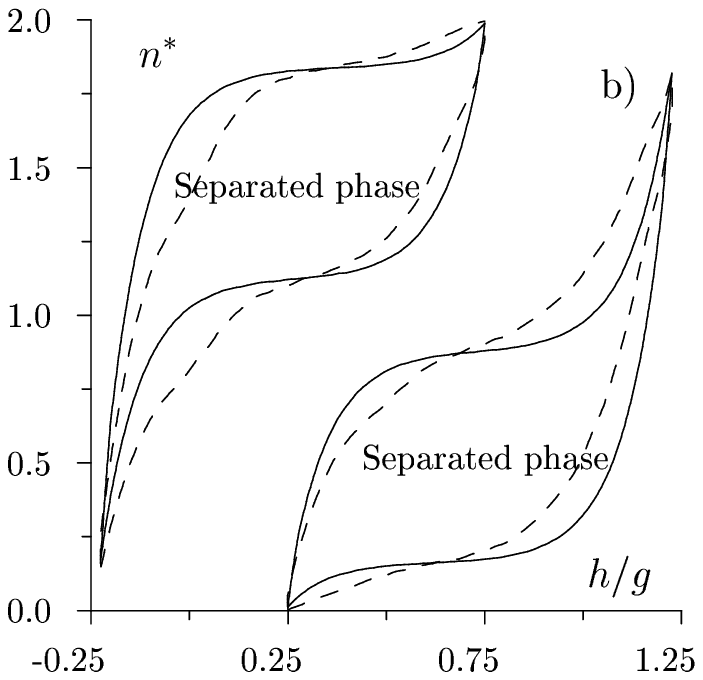}
\end{minipage}
\end{center}

\caption{Regions of phase separation at different values of interaction
$J_{\alpha\beta}$, $T/g=0.1$.
Thick lines correspond to the case  $t_{ij}=0$; dashed lines
correspond the the case
$t_{ij}\ne0$.
a) $J_{11}=g, J_{12}=0$, $t_{ij}/g=0.2$;
b) $J_{11}=J_{12}=g/2$, $t_{ij}/g=0.1$.}
\label{fig:6}
\end{figure}

Fig.\ref{fig:5}
shows that even though the solution of the system of equations
indicates that the ferroelectric phase arose (Fig.\ref{fig:5}a), the
separation takes place before that (Fig.\ref{fig:5}b). It is also indicated by a certain
concavity in the free energy behaviour
(tangent dashed lines in Fig.\ref{fig:5}c link the points
with concentration values $n_1$, $n_2$ and $n_3$, $n_4$ on which
the separation take place).
Hence, in the $n=const$ regime we can talk only about the
separation on the paraelectric and ferroelectric phases at concentration values
$n_1<n<n_2$ and $n_3<n<n_4$. The ferroelectric phase exists (in the case considered
in Fig.\ref{fig:5}) in the concentration range $n_2<n<n_3$.
Fig.\ref{fig:6}a and \ref{fig:6}b show
the boundaries of the separated phase with and without
transfer, positions of which are temperature dependent.
At $T=0$ the region of phase separations reaches the sides
of the $0\le h/g\le1$, $0\le n^*\le2$ \cite{SH98} rectangle. The transfer
influence is ambiguous. In the first case (\ref{fig:6}a) it narrows
the region of
separations, whereas in the second case (\ref{fig:6}b), the
transfer only changes the shape of the region boundaries. It should be
noted that in the presented figures, the high temperature is taken;
therefore, at chosen values of model parameters and in the absence of direct pseudospin-pseudospin interaction,
the separation does not exist.

% ---------------------

\section{Conclusions}
The mean field approximation, being the simplest one in the theory of
many-particle systems, is not trivial for the models with strong
correlations of charge carriers. In this case, the effective field is
created by transfer of electrons from site to site, not by regular
averaging of a direct interaction. In contrast to the case
of Fermi-systems, we have to solve here not only the equation for the
self-energy part, but also some extra equations for the averages of the
Hubbard operators \cite{Izyumov94_J}. In this paper we show that this approach means a
selection of diagrams with only one sum over the wavevector {\bf k} and
frequency $\omega$ in expressions for the self-energy part and averages and is a
generalization of the GRPA.
{\bf Obtained shift of lower Hubbard band corresponds to
results of the moment technique \cite{Harris,Eskes,Pothoff} in paraphase and
in the limit $U\rightarrow\infty$}.
The MFA equations for
self-consistency parameters and expressions for thermodynamic
functions allow one to analyze the equilibrium states and phase diagrams for
the systems with strong correlation.

In \cite{Letfulov98_E1} it has been shown how to obtain an equation of the dynamic mean field
approximation ($d=\infty$) for the Falicov-Kimball model by means of
summation of diagrams and selecting the single-site contributions.
Nevertheless, the question of analytical consideration of such contributions
remains in general still open. Hence,
we hope that a self-consistent approach, similar to the presented one,
will  give us a clue how to solve a single-site problem in the
infinite dimensions for the another examples of models with short-range
electron interaction.

The mean field approximation applied to the two-sublattice model with local
anharmonicity indicates a possibility of the phase separation in the
system.  It has been shown in \cite{SH98} that the direct pseudospin-pseudospin
interaction is sufficient to make a system  divide into regions with
different concentration of charge carriers. Instabilities with respect to
phase separation were found in the ferromagnetic Condo model (see for example \cite{Manchini98}),
single-sublattice pseudospin-electron model with $U=0$ and without direct
interaction between pseudospins \cite{SST99}.
From this point of view, our paper shows that the electron transfer not only
does not destroy the separation regions, but facilitates
process separation and even is able to form separated
phases by itself. This effect takes place when the chemical potential $\mu$
lies near the band edge, and the change of energy spectrum (due to ordering
of pseudospins) in the
$\mu=const$ regime moves the chemical potential into or out of the band.
One can conclude that the electron
(hole) transfer  in the pseudospin-electron model creates an effective
interaction between pseudospins. It is qualitatively different from the standard
one $J_{ij}S_i^z S_j^z$ \cite{SSD95}, but its role is similar.

The problem of applicability of the approach  proposed in this paper,
as well as of the GRPA itself, requires a thorough analysis.
The approach can be improved at the expense of complication of
loop diagrams and
vertices. Here the magnon Green function $g^{43}({\bf k},\omega)$ arise.
Taking into account the magnon contributions is especially important at
$U\ne\infty$, since without them the system of equations for the averages
of the Hubbard operators  becomes dependent on the priority rules. These
problems are the subject of our subsequent studies.

%It should be noted that the considered approximation has some
%drawbacks:  neglected magnon Green's functions $g^{43}({\bf k},\omega)$
%can be essential even in the paraphase. Moreover, without taking them
%into account the system of equations for the averages of the Hubbard
%operators at $U\ne\infty$ becomes asymmetric. Taking into account the
%magnon correlations in the paraphase is a problem we are working on
%now.


\begin{thebibliography}{19}

\bibitem{Izyumov94_J}
   Izyumov~Yu.A., Letfulov~B.M. and Shipitsyn~E.V.,
   A mean-field-type approximation for the ($t-J$) model.
   // J. Phys.: Condens. Matter, 1994, vol.~6, p.~5137-5154.

\bibitem{Hubbard63}
      Hubbard~J., Proc. R. Soc. London, 1963, Ser~A vol.~276, p.~238.

\bibitem{Abrikosov75}
   Abrikosov~A.A., Gorkov~L.P. and Dzyaloshinskii~I.E.,
   Methods of quantum field theory in statistical physics.
   // 1975, New York: Dover.

\bibitem{Izyumov92_I}
   Izyumov~Yu.A., Letfulov~B.M., Shipitsyn~E.V. and Chao~K.A.,
   A theory of ferromagnetism in the Hubbard model with infinite Coulomb
   interaction.
   // Int. Journ. Mod. Phys. B, 1992, vol.~6, No.~21, p.~3479-3514.

\bibitem{Izyumov92_P}
   Izyumov~Yu.A., Letfulov~B.M., and Shipitsyn~E.V., Bartkowiak~M. and Chao~K.A.,
   Theory of strongly correlated electron systems on the basis of
   diagrammatic technique for Hubbard operators.
   // Phys. Rev. B, 1992, vol.~46, No.~24, p.~15697-15711.

\bibitem{Izyumov92_J}
   Izyumov~Yu.A., Letfulov~B.M. and Shipitsyn~E.V.,
   Generalized random-phase approximation in the theory of strongly
   correlated systems.
   // J. Phys.: Condens. Matter, 1992, vol.~4, p.~9955-9970.

\bibitem{SS93} Stasyuk~I.V., Shvaika~A.M.
      Dielectric properties and electron spectrum of the M\"uller model in
      the {HTSC} theory.  // Acta Physica Polonica~A, 1993, vol. 84, No 2,
      p.~293-313.

\bibitem{SS94}  Stasyuk~I.V., Shvaika~A.M.
   Dielectric instability and local anharmonic model in the theory of high--T$_{c}$ superconductivity.
   // Physica C, 1994, vol.~235-240, p.~2173-2174.

\bibitem{SSD95} Stasyuk~I.V.,
      Shvaika~A.M. and Danyliv~O.D.  Dielectric instability and charge ordering in
      the local anharmonic model of high $T_c$ superconductors. // Molecular
      Physics Reports, 1995, vol.~9, p.~61-75.

\bibitem{Letfulov98_E1}
   Letfulov~B.M.,
   Strongly correlated Falicov-Kimball model in infinite dimensions
   // Eur. Phys. J. B, 1998, vol.~4, p.~447-457.

\bibitem{Letfulov98_E2}
   Letfulov~B.M.,
   A theory of double exchange in infinite dimensions.
   // Eur. Phys. J. B, 1998, vol.~4, p.~195-203.

\bibitem{SS99}
   Stasyuk~I.V., Shvaika~A.M.
   Pseudospin-electron model in large dimensions.
   // J. Phys. Studies, 1999, vol.~3, No.~2, p.~177-183.

\bibitem{Ruani94} Ruani~G., Taliani~C., Muccini~M., Conder~K.,
      Kaldis~E., Keller~H., Zech~D., Muller~K.A.  Apex anharmonicity observed by
      Raman scattering in ${}^{18}$O substituted YBa$_2$Cu$_3$O$_{6+x}$. //
      Physica C, 1994, vol.  226, p. 101-105.

\bibitem{Mihailovic90}
      Mihailovic~D. and Heeger~A.J., Pyroelectric and piezoelectric effects in
      single crystals of {YBa$_2$Cu$_3$O$_{7-\delta}$}. // Solid State Comm.,
      1990, vol.~75, p.~319-323.

\bibitem{Grachev97}
      Grachev~A.I. and Pleshkov~I.V., Pyroelectric voltage in YBCO thin films.//
      Solid State Comm., 1997, vol.~101, No.~7, p.~507-512.

\bibitem{Mustre92} Mustre de Leon~J., Conradson~S.D.,
      Batistic'~I., Bishop~A.R., Ra\-is\-trick~I.D., Aronson~M.C., and Garzon~F.H.
      Axial oxygen-centered lattice instabilities in YBa$_2$Cu$_3$O$_{7}$:
      An application of the analysis of extended x-ray-absorbtion fine
      structure in anharmonic systems. // Phys.Rev.B, 1992, vol. 45,
      p.2447-2457.

\bibitem{Conradson89}
      Conradson~S.D., Raistrick~I.D.  The  axial  oxygen  atom  and
      superconductivity in YBa$_2$Cu$_3$O$_7$.  // Science, 1989, vol. 243,
      No 4896, p.~1340.

\bibitem{r4} M\"uller~K.A.  // Phase transitions,
      1988 (Special issue).

\bibitem{Stas74} Stasyuk~I.V., Slobodyan~P.M.
  The diagrammatic technique for Hubbard operators.
   // Theor.Math.Phys. USSR, 1974, vol.19, p.~616
   [TMF., 1974, vol.~19,
   No~3, p.~\mbox{423-428} {\it (in Russian)}], 
   Westwenski~B. // Phys.Lett.A, 1973, vol.44, p.~27.

\bibitem{Izyumov89} Izyumov~Yu.A., Skryabin~Yu.N.
   Statistical Mechanics of Magnetically Ordered Systems
   // 1989, New York: Consultants Bureau.

\bibitem{DanStas98} Danyliv~O.D.
      Phase transitions in the two-sublattice pseudospin-electron model
      of high temperature superconducting systems. // Physica C, 1998, vol.~309, p.~303-314.

\bibitem{SH98} Stasyuk~I.V., Havrylyuk~Yu.
   Phase separation in pseudospin-electron model with direct interaction
   between pseudospins.
   // Cond. Mat. Phys., 1999, vol.~2, No~3 (19).

\bibitem{Harris} Harris~A.B. and Lange~R.V.
   Single-particle excitations in narrow energy bands.
   // Phys.Rev., 1967, vol.~157, No~2, p.295-314.

\bibitem{Eskes} Eskes~H., Oles'~A.M., Meinders~M.B.J. and Stephan~W.
   Spectral properties of Hubbard bands.
   // Phys.Rev.B, 1994, vol.~50, No~24, p.17980-18002.

\bibitem{Pothoff} Pothoff~M., Herrmann~T., Wegner~T., and Nolting~W.
   The momentum sum rule and its consequences for ferromagnetism.
   // Phys.Stat.Sol.(b), 1998, vol.~210, p.199-227.

\bibitem{Manchini98}
   Moreo~A., Avella~A., and Cuoco~M.
   Numerical studies of strongly correlated electron systems. In
   ``Lectures on the physics of highly correlated electron systems''. Ed. by
   Manchini~F. // 1998, New York: Woodbury.

\bibitem{SST99}
   Stasyuk~I.V., Shvaika~A.M., Tabunshchyk~K.V.
   Thermodynamics of pseudospin-electron model without correlations.
   // Cond. Mat. Phys., 1999, vol.~2, No~1 (17), p.~109-132.

\end{thebibliography}
\end{document}